# Optimal estimates and joint measurement uncertainty relations[*]

Michael J.W. Hall
Theoretical Physics, IAS, Canberra ACT 0200, Australia
*mjh105@rsphysse.anu.edu.au*

## Abstract
Often, one would like to determine some observable *A*, but can only measure some (hopefully related) observable *M*. This can arise, for example, in quantum eavesdropping, or when the research lab budget isn't large enough for a 100% efficient photodetector. It also arises whenever one tries to *jointly* determine two complementary observables *A* and *B*, via some measurement *M*.

This raises three natural questions:
(i) what is the best possible estimate of *A* from *M* ?
(ii) how 'noisy' is such an estimate ?
(ii) are there any universally valid uncertainty relations for joint estimates ?

Quite general answers, and applications to heterodyne detection and EPR joint measurements, are briefly reviewed.

# 1. Thought experiment

Would like to determine observable *A*, but can only measure observable *M*.

*What is **the best possible estimate** of A from the measurement result M=m ?*

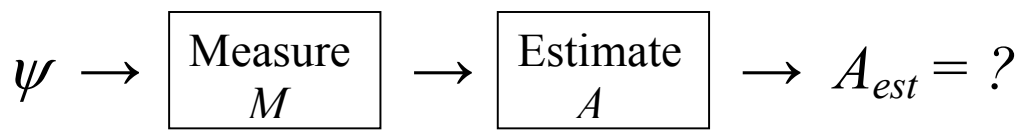

$\psi \rightarrow \boxed{\text{Measure } M} \rightarrow \boxed{\text{Estimate } A} \rightarrow A_{est} = ?$

This question arises, for example in

- ❖ Quantum eavesdropping

- ❖ Joint measurements of two quantum observables *A* and *B* (eg, position and momentum)

- ❖ Classical estimation theory

- ❖ Non-ideal lab equipment (?!)

## 2. How good is a given estimate ?

If *f(m)* denotes the estimate of *A* from measurement result *M=m*, then the estimate is equivalent to measuring the observable

$$A_f = f(M) = \Sigma_m \, f(m) \, |m\rangle \langle m| \, .$$

Hence can decompose any estimate as[1,2]

$$A_f = A + N_f \, ,$$

i.e.,

$$\boxed{\textit{estimate = signal + noise} \, ,}$$

where $N_f$ is the **noise operator** associated with the estimate.

The **inaccuracy in the estimate** is defined to be the **rms noise**:

$$\varepsilon(A_f)^2 := \langle N_f^{\,2} \rangle,$$

which vanishes for a *perfect* estimate.

# 3. Optimal estimate ≡ smallest noise

It may be shown that *the noise has the fundamental lower bound* [3-5]

$$\varepsilon(A_f)^2 \geq \sum_m \left| \text{Im} \frac{\langle m|A|\psi\rangle}{\langle m|\psi\rangle} \right|^2, \quad (1)$$

and hence that [5]

*incompatibility* ⇒ *noise*

(i.e., $\varepsilon(A_f) > 0$ for non-commuting $A$ and $M$).

The lower bound in (1) corresponds to the *optimal estimate* of $A$, which is given, for measurement result $M=m$ on state $\psi$, by [3-5]

$$A_{opt} = \sum_m \text{Re} \frac{\langle m|A|\psi\rangle}{\langle m|\psi\rangle} |m\rangle\langle m|.$$

One finds a *spread vs noise* tradeoff [5]:

$$(\Delta A_{opt})^2 + \varepsilon(A_{opt})^2 = (\Delta A)^2,$$

i.e., a *geometric uncertainty relation*.

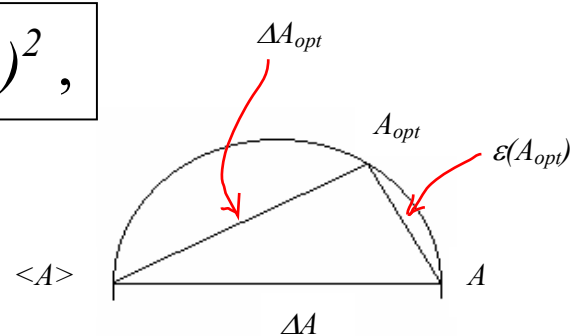

# 4. Aside: examples of optimal estimates

**_Momentum:_** Writing $\psi = R e^{iS/\hbar}$, the optimal estimate of momentum, from a position measurement result $X=x$, follows as

$$P_{opt}(x) = \nabla S.$$

This estimate achieves the lower bound in (1), which can be rewritten as an **_exact uncertainty relation_** [3]:

$$\delta X \, \varepsilon(P_{opt}) = \hbar/2.$$

> $\delta X$ is a classical measure of uncertainty called the Fisher length

This implies, and is far stronger than, the Heisenberg uncertainty relation $\Delta X \Delta P \geq \hbar/2$.

**_Energy:_** The optimal estimate for energy is

$$E_{opt}(x) = |\nabla S|^2/(2m) + V + Q,$$

where $Q = -\hbar^2 \nabla^2 R/(8mR)$ is the so-called quantum potential.

# 5. Joint measurements

**All *measurements are joint measurements* !**

*Why?* - the information gained from any measurement $M$ can always be used to make estimates of any two observables $A$ and $B$, via

$$A_f = f(M) \quad \text{and} \quad B_g = g(M) .$$

## *Example: Heterodyne detection*[5]

The statistics of heterodyne detection are given by the Husimi Q-function

$$Q(\alpha) = |\langle \alpha | \psi \rangle|^2 / \pi . \quad (\alpha = \alpha_1 + i\alpha_2)$$

A *standard* joint estimate of the quadratures $X=(a+a^\dagger)/2$, $Y=(a-a^\dagger)/2i$, is then:

$$X_{est} = \alpha_1, \quad Y_{est} = \alpha_2, \quad \Rightarrow \quad \boxed{\Delta X_{est} \Delta Y_{est} \geq 1/2 .}$$

The *optimal* joint estimate is *4 times better!*:

$$X_{opt} = \alpha_1 + \tfrac{1}{4} \partial_1 \ln Q, \quad Y_{opt} = \alpha_2 + \tfrac{1}{4} \partial_2 \ln Q ,$$

$$\Rightarrow \quad \boxed{\Delta X_{opt} \Delta Y_{opt} \geq 1/8 .}$$

# 6. A *universal* joint measurement uncertainty relation

Let $A_f$ and $B_g$ denote **any** two estimates of observables $A$ and $B$, from some measurement $M$ on state $\psi$. One then has the ***joint uncertainty relation*** [5,6]

$$\Delta A_f\, \varepsilon(B_g) + \varepsilon(A_f)\, \Delta B_g + \varepsilon(A_f)\, \varepsilon(B_g) \geq \tfrac{1}{2} |\langle [A,B] \rangle| . \quad (2)$$

This relation is the long-looked for universal quantification of complementarity:

*For two incompatible observables $A$ and $B$, there is no joint estimate having both zero spread and zero noise.*

## *Special case: unbiased estimates*

If a measurement $M$ yields estimates of $A$ and $B$ which are on average equal to $\langle A \rangle$ and $\langle B \rangle$, for *all* states $\psi$, then[1,2,5,6]

$$\varepsilon(A_f)\, \varepsilon(B_g) \geq \tfrac{1}{2} |\langle [A,B] \rangle| ,$$

i.e., **unbiased** *estimates of incompatible observables cannot be arbitrarily accurate*.

# 7. Example: EPR estimates and continuous variable teleportation

In continuous variable teleportation[7,8], Alice and Bob ideally share a perfect EPR state:
$$\psi_{EPR}(x,x') = \delta(x-x'-a)\, e^{ib(x+x')/2\hbar}.$$

But such states are unphysical: in practice they must use the *approximate* EPR state

$$\psi = K \exp[\, -(x-x'-a)^2/4\sigma^2 - \tau^2(x+x')^2/4\hbar^2 ]\, e^{ib(x+x')/2\hbar},$$

with the *almost* perfect correlations:

$$\langle X-X' \rangle = a, \quad \mathrm{Var}(X-X') = \sigma^2 \ll 1,$$
$$\langle P+P' \rangle = b, \quad \mathrm{Var}(P+P') = \tau^2 \ll 1.$$

Now, if Alice transmits a measurement result $P=p$ to Bob, then **what is the best estimate Bob can make for $P'$?**

$$P'_{est} = b-p \quad \boldsymbol{\times} \qquad P'_{opt} = \frac{\hbar^2(b-p) + \sigma^2\tau^2 p}{\hbar^2 + \sigma^2\tau^2} \quad \checkmark$$

$$\boxed{\varepsilon(P'_{opt})/\varepsilon(P'_{est}) = (1+\sigma^2\tau^2/\hbar^2)^{-1/2} < 1}$$

$\Rightarrow$ ***optimal estimates can improve the teleportation protocol*** - and achieve the fundamental lower bound in (2).

# 8. Generalisations: POMs and density operators

All of the main results above can be generalised to the case where

- ❖ the measurement $M$ is described by a *probability operator measure* (POM), i.e., by a set of positive operators $\{M_m\}$ with $\Sigma_m M_m = 1$ (eg, $M_m = |m\rangle\langle m|$).

- ❖ the state of the system prior to measurement is described by a density operator $\rho$ (eg, $\rho = |\psi\rangle\langle\psi|$).

***The geometric and joint measurement uncertainty relations remain unchanged***, and the lower bound (1) for noise, and the formula for the optimal estimate take the respective forms[5]

$$\varepsilon(A_f)^2 \geq \Sigma_m \frac{|tr[\rho(AM_m - M_mA)]|^2}{4\,tr[\rho M_m]},$$

$$A_{opt} = \Sigma_m \frac{tr[\rho(AM_m + M_mA)]}{2\,tr[\rho M_m]}\,|m\rangle\langle m|.$$

# 9. Summary

Any estimate of an observable *A* from some measurement *M* can be decomposed as
*estimate = signal + noise* .

The noise cannot vanish if *A* and *M* are incompatible:
*incompatibility* ⇒ *noise* .

There is a ***geometric uncertainty relation*** for optimal estimates, reflecting a ***fundamental trade-off between the spread and the noise of an optimal estimate***:
$$(\Delta A_{opt})^2 + \varepsilon(A_{opt})^2 = (\Delta A)^2$$

There is a ***universal joint measurement uncertainty relation***, valid for the estimates of any two observables *A* and *B* from any measurement process:
$$\Delta A_f \, \varepsilon(B_g) + \varepsilon(A_f) \, \Delta B_g + \varepsilon(A_f) \, \varepsilon(B_g) \geq \tfrac{1}{2} \, |\langle [A,B] \rangle|$$

# **References**

(NB: material for this poster is largely drawn from Ref. 5, where many more details and examples may be found)

1. E. Arthurs and J.L. Kelly, *Bell Syst. Tech. J.* **44** (1965) 725

2. D.M. Appleby, *Int. J. Theor. Phys.* **37** (1998) 1491

3. M.J.W. Hall, *Phys. Rev. A* **64** (2001) 052103

4. L.M. Johansen, *Phys. Lett. A* **322** (2004) 298

5. M.J.W. Hall, *Phys. Rev. A* **69** (2004) 052113

6. M. Ozawa, *Phys. Lett. A* **320** (2004) 367

7. W.P. Bowen *et al.*, *Phys. Rev. A* **67**, (2003) 032302

8. C.M. Caves & K. Wodkiewicz, *eprint* quant-ph/0409063